\pdfoutput=1

\documentclass[twocolumn,amsmath,amssymb]{revtex4}
\usepackage{graphicx}

\begin{document}

\title{Electroluminescence from single nanowires by tunnel injection:\\ an experimental study}
\author{Mariano A. Zimmler$^1$, Jiming Bao$^1$, Ilan Shalish$^1$, Wei Yi$^1$, Joonah Yoon$^{1,2}$,\\ Venkatesh Narayanamurti$^1$ and Federico Capasso$^1$}\email{capasso@seas.harvard.edu}
\affiliation{$^1$Harvard University, Cambridge, MA 02138}
\affiliation{$^2$Massachusetts Institute of Technology, Cambridge, MA 02139}

\date{\today}

\begin{abstract}

We present a hybrid light-emitting diode structure composed of an n-type gallium nitride nanowire on a p-type silicon substrate in which current is injected along the length of the nanowire. The device emits ultraviolet light under both bias polarities. Tunnel-injection of holes from the p-type substrate (under forward bias) and from the metal (under reverse bias) through thin native oxide barriers consistently explains the observed electroluminescence behaviour. This work shows that the standard p--n junction model is generally not applicable to this kind of device structure.

\end{abstract}

\maketitle

\section{Introduction}

Since its birth in 1907 \cite{Round1907}, the light-emitting diode (LED) has evolved into a mature technology, providing electrically-generated light in a myriad of applications including large-scale displays, signage, and white light sources \cite{Schubert2006}. However, many future applications, such as integrated photonics, rely on the miniaturization and integration of LEDs of different colours on a single chip. A viable avenue to achieve this is the use of semiconductor nanowires. Nanowires of the group III nitrides are especially attractive candidates given that their emission wavelength can be tuned from the infrared---indium nitride---to ultraviolet---gallium nitride---through variations in alloy composition. Gallium nitride (GaN) is the most mature of them, i.e., the backbone of the family, which serves as the base for most of the published applications to date \cite{Nakamura2000}. Recent work has demonstrated electroluminescence (EL) from GaN nanowires in a cross-wire geometry \cite{Huang2005}, in single nanorod p--n junctions \cite{Kim2003}, as well as in core/multishell heterostructures \cite{Qian2005}. However, in all of these approaches carrier injection is generally limited to a small fraction of the 
\begin{figure}[bp]
   \centering
   \includegraphics[width=0.4\textwidth]{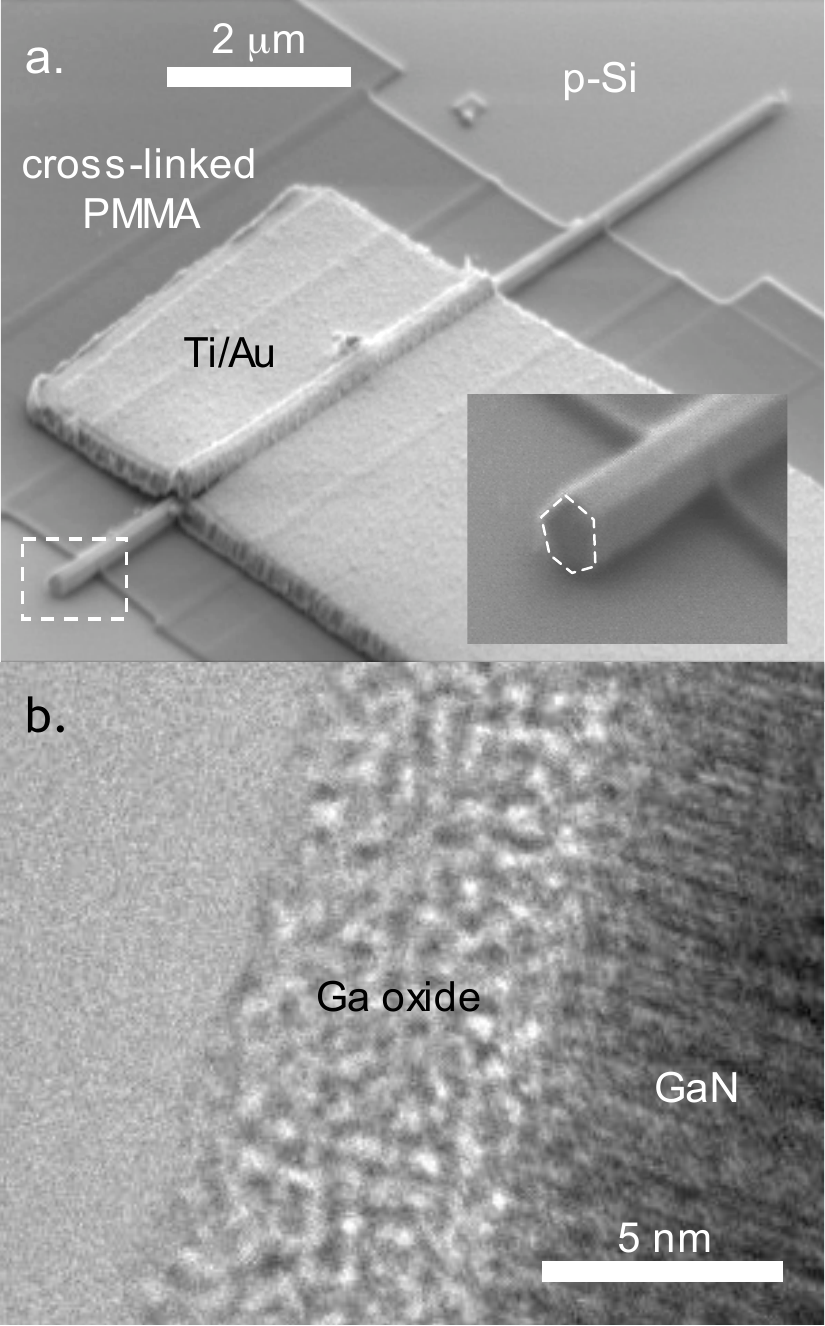}
   \caption{(a) SEM image of a typical n-GaN nanowire/p-Si light-emitting diode. Inset: Zoomed-in image of the nanowire end. (b) High-resolution TEM image of a GaN nanowire exhibiting a $\sim$5 nm shell of native oxide.}
   \label{fig:sem}
\end{figure}
semiconductor material, which limits the output power. A more efficient architecture would make use of the entire length of the nanowire cavity for current injection which, together with proper feedback \cite{Barrelet2006}, could also be suitable for nanowire lasing. In this paper, we report electroluminescence from a hybrid structure composed of a n-type GaN nanowire in contact with a p-type Si substrate, where the current is injected along the length of the nanowire. This scheme, first demonstrated with cadmium sulfide nanowires \cite{Duan2003} and later implemented in zinc oxide \cite{Bao2006}, is attractive for several reasons. First of all, it does not require complex nanowire growth techniques involving either axial or radial dopant modulation. Second, it can be applied to materials that are poor amphoteric semiconductors, i.e. semiconductors that are difficult to obtain in both n- and p-type. Finally, device assembly does not require complicated fabrication steps, as is the case with core/multishell heterostructures (e.g. selective etching of the nanowire shell to contact the core). As we show, our structure emits ultraviolet EL from the GaN band edge under both bias polarities, suggesting that light emission is mediated by tunnel injection of carriers through thin native oxide barriers \cite{Jaklevic1963, Fischer1963}. This work therefore provides strong evidence that the standard p--n junction model is generally not applicable to this kind of device structure.

\section{Experimental methods}

Single-crystal wurtzite GaN nanowires were grown on \emph{c}-plane sapphire substrates by hydride vapour phase epitaxy, using nickel-gold as a catalyst. Typical nanowires were several microns long and $\sim$150 nm in diameter. Details of the growth were given elsewhere \cite{Seryogin2005}. The nanowires were removed from the growth substrate using ultrasonic agitation in ethanol and randomly dispersed on the device substrate. Conductivity measurements on nanowire channel field effect transistors show typical n-type carrier concentrations in excess of $10^{18}$ cm$^{-3}$. This value is commonly observed and reported for GaN nanowires \cite{Kim2003,Huang2002}.

LEDs were assembled by placing the nanowires on a heavily doped p-type silicon (p-Si) substrate (hole concentration $\sim$ $3 \times 10^{19}$ cm$^{-3}$), and subsequently defining a metallic top contact as in \cite{Bao2006}. Figure \ref{fig:sem}(a) shows a scanning electron microscope (SEM) image of a typical device. A thin layer of lithographically defined cross-linked PMMA separates the substrate from the metallic pads on the nanowire sides. The inset highlights the hexagonal cross-section of the nanowire, which suggests that its bottom facet provides a large flat interface to the silicon substrate. In the following discussion, we shall refer to positive (or forward) bias as the potential of the p-Si substrate with respect to the metallic contact, and therefore to hole injection into the nanowire from the p-Si substrate. We fabricated and characterized 7 n-GaN nanowire/p-Si devices. All data presented herein were collected from the same representative device. Variations among different devices will be discussed in the relevant context.

\section{Results and discussion}

\begin{figure}[tbp]
   \centering
   \includegraphics[width=0.42\textwidth]{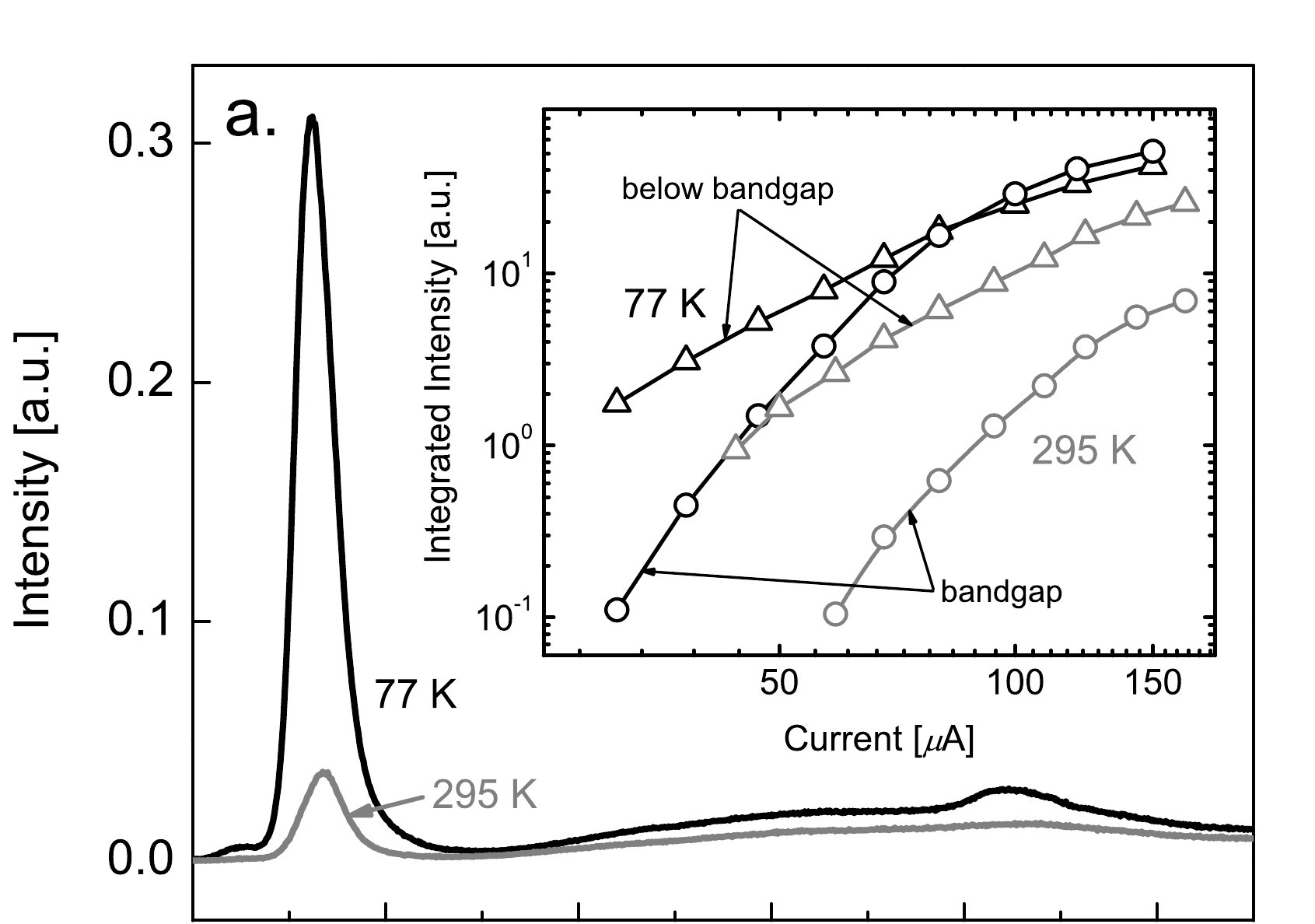}
   \includegraphics[width=0.42\textwidth]{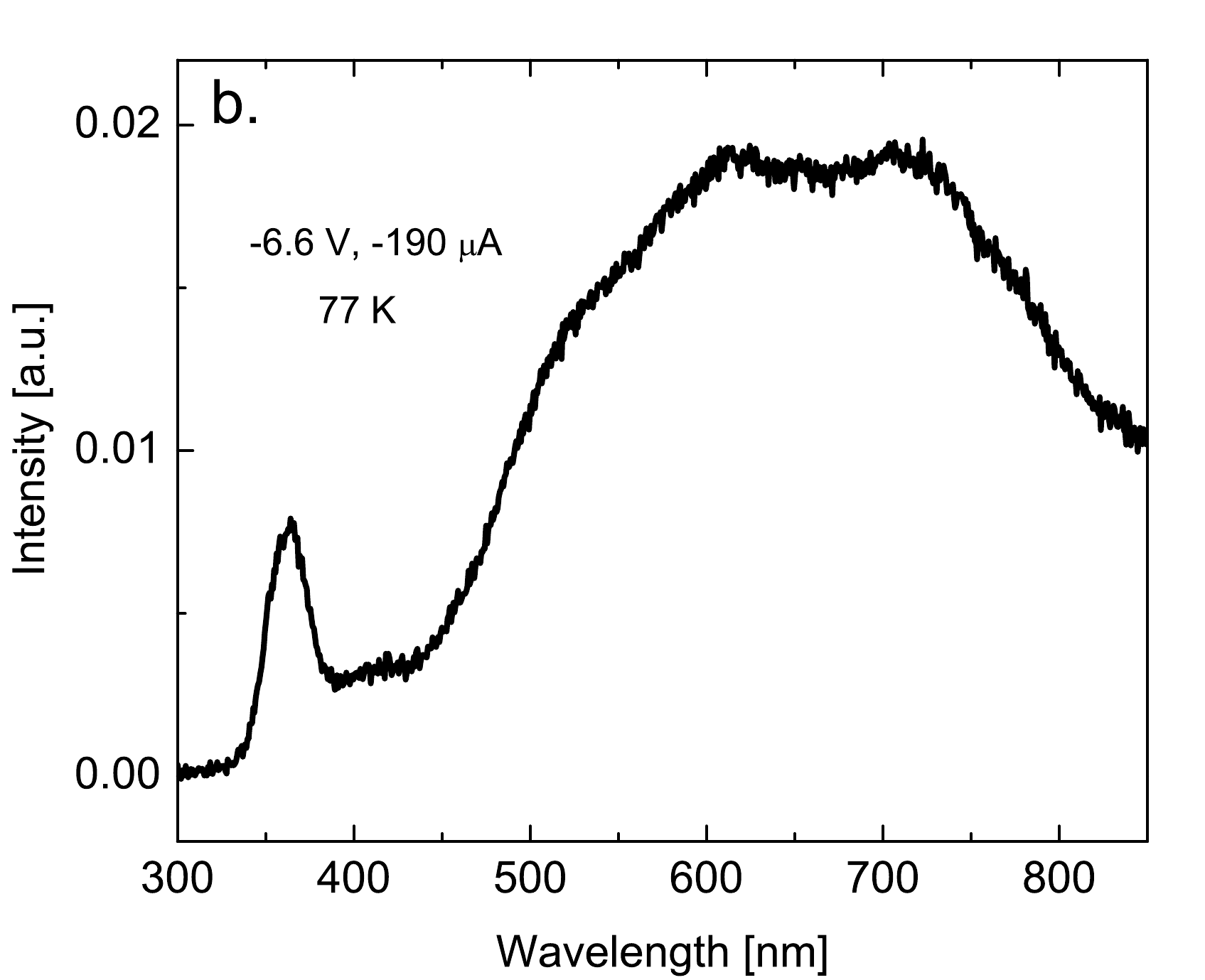}
   \includegraphics[width=0.42\textwidth]{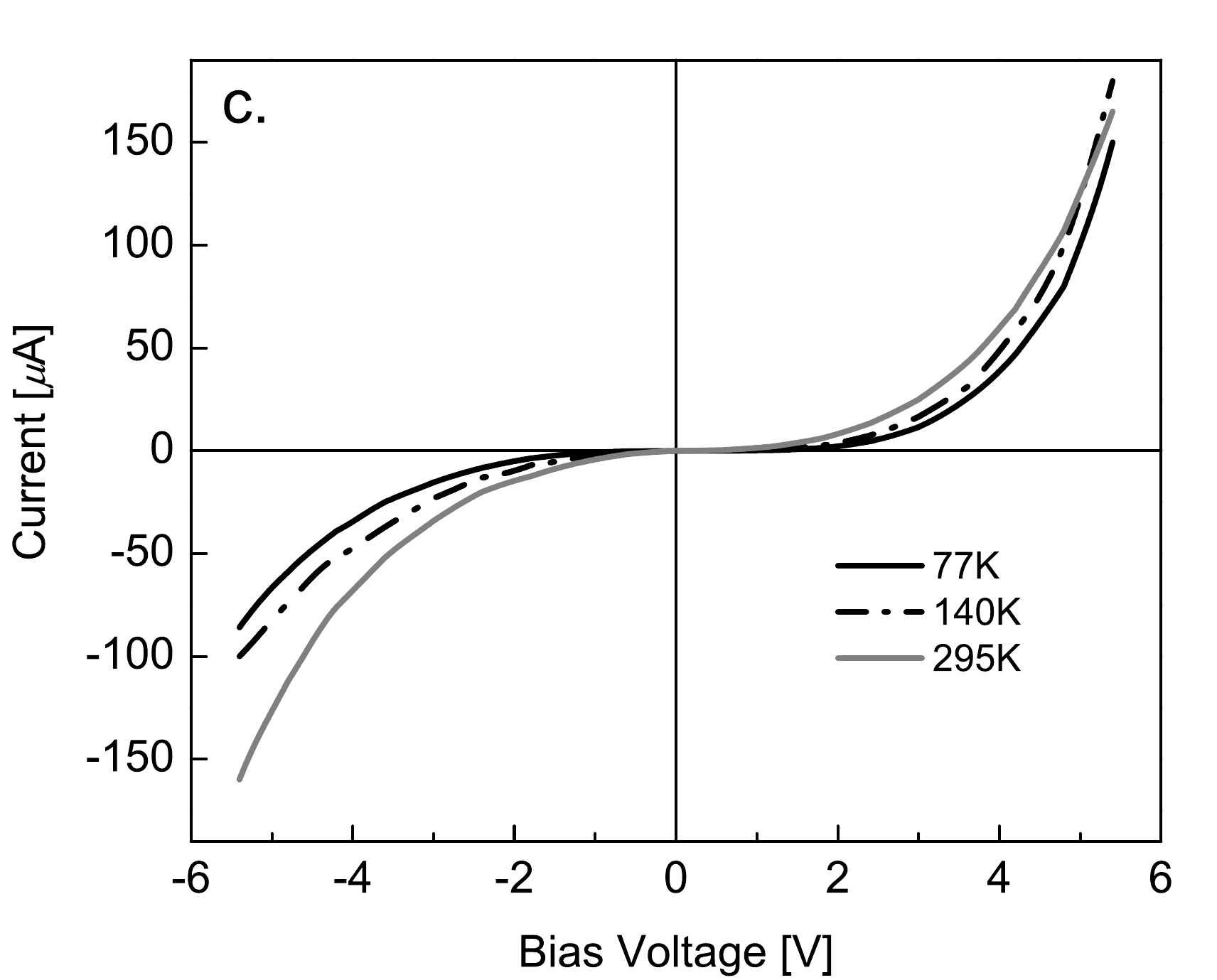}
   \caption{(a) Forward bias electroluminescence (EL) spectra of a typical device at 77 K (black) and at 295 K (grey). Both traces were obtained with the same applied bias of $+5.4$ V and currents of 150 $\mu$A at 77 K and 165 $\mu$A at room temperature. The small peak centred near 720 nm on the 77 K trace corresponds to the grating's second order diffraction of the band edge peak seen at 360 nm. Inset: plot of integrated intensity versus current at 77 K (black) and 295 K (grey). The integrated bandgap EL (open circles) was obtained by adding the measured counts in the range from 300 nm to 450 nm. The below bandgap luminescence (open triangles) was integrated in the range from 450 nm to 850 nm. (b) Reverse bias EL spectrum for a typical n-GaN nanowire/p-Si device. The a.u. scale is the same in (a) and (b). (c) Current versus voltage characteristics measured at 77 K (black), 140 K (dash--dot line) and 295 K (grey).}
   \label{fig:EL}
\end{figure}

\begin{figure*}[tbp]
   \centering
   \includegraphics[width=0.85\textwidth]{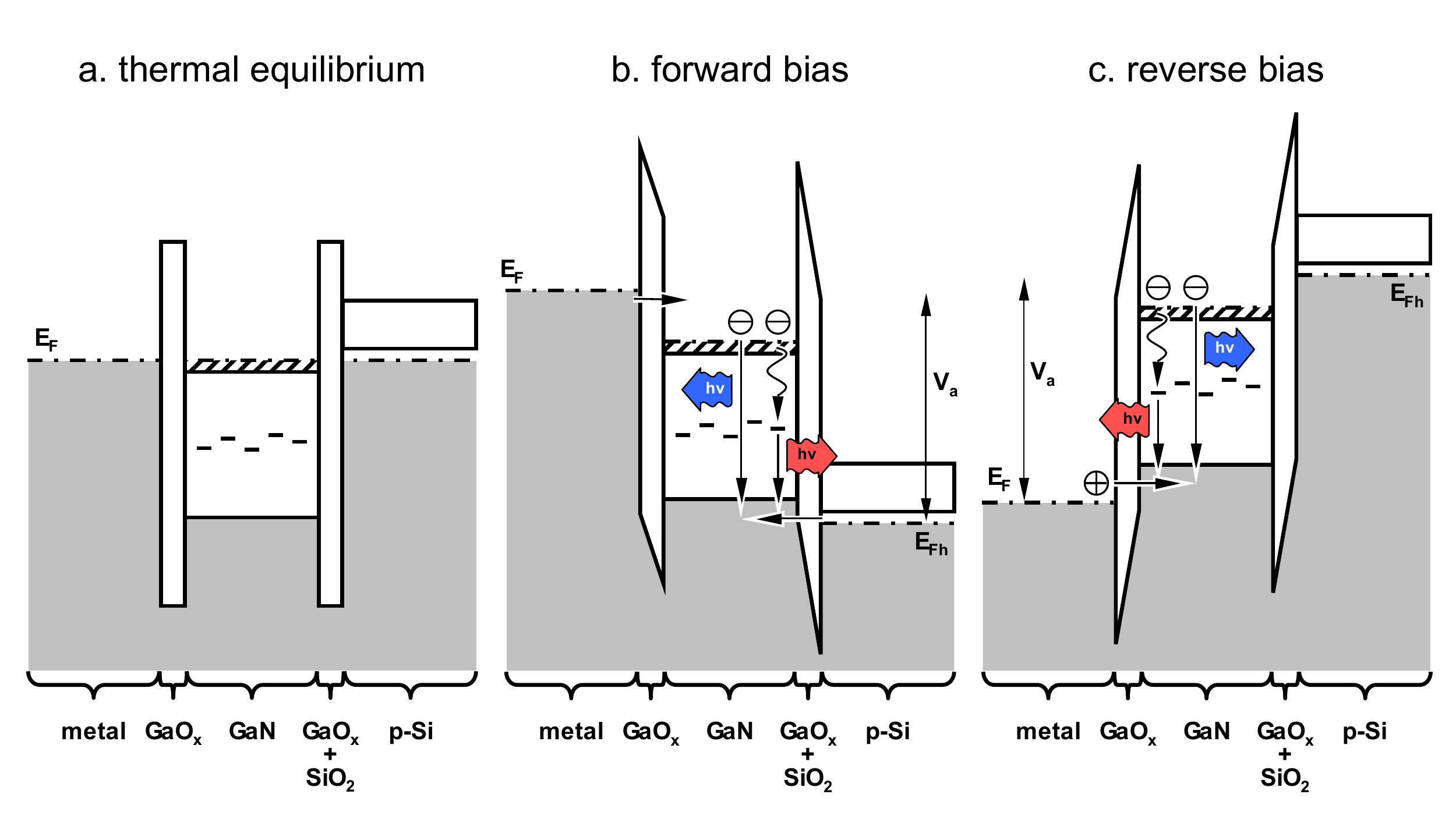}
   \caption{Schematic band diagram of a n-GaN/p-Si hybrid electroluminescent device (a) in thermal equilibrium, (b) under forward bias and (c) under reverse bias.}
   \label{fig:bd}
\end{figure*}

Under moderate forward bias ( $V < 2E_{\mathrm{g}}/e$, where $E_{\mathrm{g}} \sim 3.5$ eV is the bandgap of GaN \cite{Madelung2003}), our LEDs emit mostly near band-edge luminescence. EL spectra obtained under a forward bias of 5.4 V, at liquid nitrogen temperature (black) and at room temperature (grey), are shown in figure \ref{fig:EL}(a). In both cases, the spectrum consists of a dominant ultraviolet peak and a broad sub-bandgap peak. At 77 K, the ultraviolet emission is more intense, peaks at 361 nm, and has a full width at half maximum of 23.5 nm. This wavelength corresponds to an energy of 3.43 eV, which is close to the bandgap of GaN at 77 K ($E_{\mathrm{g}} \sim 3.5$ eV) \cite{Madelung2003}. Quantization effects are negligible in our nanowires. The inset in figure \ref{fig:EL}(a) shows the current dependence of the band-edge EL (circles) and of the broad sub-bandgap emission (triangles). The estimated collected power at 77 K, for bandgap luminescence, at a current of $\sim$165 $\mu$A, is $\sim$0.2 pW.

Remarkably, these nanowire devices also show EL under reverse bias \cite{Bao2006}. Figure \ref{fig:EL}(b) shows a typical spectrum collected under a reverse bias of $-6.6$ V ($I \sim 190$ $\mu$A) at 77 K (see figure \ref{fig:EL}(a)). Compared with forward bias, the overall EL intensity is weaker and the near bandgap luminescence no longer dominates the spectrum. Figure \ref{fig:EL}(c) shows the current--voltage characteristics, obtained at three different temperatures (77 K, 140 K, and 295 K).

Light emission in a reverse-biased p--n junction is commonly attributed to the recombination of electron--hole pairs created by impact ionization \cite{Pankove1971}. This model, however, is not applicable to our devices. Avalanche multiplication is characterized by broad-band emission at energies substantially higher than the bandgap (resulting from hot-carrier recombination) and negligible band-edge emission, and consequently requires significantly larger applied voltage than under forward bias. In contrast, the near band-edge emission obtained under reverse bias from our devices is clearly visible, and the applied voltage is only slightly larger than under forward bias.

To understand the behaviour of these devices, it is necessary to recognize that p--n junctions formed by bringing two semiconductors into contact (the n-type nanowire and the p-type substrate, in our case) are far from ideal, compared to standard epitaxially grown p--n junctions, leading to larger turn-on voltage and non-ideal $I\mathrm{-}V$ characteristics. First, in contrast to the latter, covalent chemical bonds between the two semiconductors are not formed. Rather, the interface region arises from a mechanical contact via van der Waals forces. Second, surfaces of semiconductors develop thin oxide layers upon exposure to ambient air. In fact, silicon substrates generally have a layer of a native oxide approximately 1 nm thick. GaN bulk \cite{Watkins1999} and, in particular, GaN nanowires \cite{Tang2003}, are also prone to oxidation, with oxide layers typically several nanometers thick for nanowires with large diameters ($>$40 nm) \cite{Tang2003}. Figure \ref{fig:sem}(b) shows a high-resolution transmission electron microscope (TEM) image of a nanowire from the same growth run as those used in our devices, which exhibits a $\sim$5 nm oxide shell surrounding the nanowire core---as determined by energy-dispersive x-ray spectroscopy (EDS).\footnote{It should be noted that while TEM images of the nanowires in \cite{Seryogin2005} do not present native oxide, our experience shows that nanowires that are imaged a few days after growth typically do indeed exhibit a thin layer of native oxide.} We should note that EDS is not entirely conclusive in the identification of light compounds such as oxygen or carbon, so that, in principle, the coating on the nanowire could also be carbon contamination. This possibility can be ruled out by a twofold argument: first, in light of the work in \cite{Tang2003}, which demonstrates that a several nanometre thick oxide is quite common for large diameter GaN nanowires; second, by noting that a carbon layer would be very conductive compared to an oxide, which is clearly not supported by the measured $I-V$ characteristics. The above discussion therefore implies that both the nanowire/substrate and the nanowire/metal junctions generally have an intermediate thin oxide barrier.

Figure \ref{fig:bd}(a) shows a schematic diagram of the thermal equilibrium band diagram of the nanowire device shown along the radial direction perpendicular to the substrate. Thin oxide barriers both between the nanowire and the substrate and between the nanowire and the metal have been included. Under an applied bias, most of the voltage drops across the oxide layers, since both the silicon substrate and the GaN nanowire are heavily doped and therefore highly conducting.

Under forward bias (figure \ref{fig:bd}(b)), due to the presence of the interface oxide, it is possible for holes in the valence band of p-Si to tunnel into that of GaN, and then radiatively recombine with electrons in the conduction band, generating band-edge luminescence. This occurs when the potential energy drop between the GaN nanowire and the silicon substrate exceeds 3.5 eV (i.e., the bandgap of GaN). For the device in figure \ref{fig:EL}, the threshold voltage for near bandgap luminescence is $\sim$4.5 V. This excess voltage can be ascribed to the ohmic drop across the series resistance of the two oxide layers and is expected to be somewhat sample dependent, as observed experimentally. Similarly, the broad band sub-bandgap emission observed in figure \ref{fig:EL}(a) can be understood as resulting from the radiative recombination of electrons from filled deep levels near the GaN/Si interface with holes tunnelling from p-Si. This process will be accompanied by a concomitant nonradiative capture by the deep centre of a conduction band electron. The alternative process of luminescence from the GaN conduction band to a deep level that has captured a hole is also possible, depending on the nature of the centre.

EL under reverse bias can be understood with similar arguments (figure \ref{fig:bd}(c)). In this case, hole injection into the nanowire \emph{from the metal} becomes possible when the potential energy drop between the nanowire and the metal is approximately 3.5 eV. In other words, electrons from the valence band of GaN can tunnel out into the metal when the Fermi level of the latter drops below the valence band edge in GaN. The resulting hole in the GaN nanowire can recombine radiatively with an electron in the conduction band producing near band-edge luminescence. Note that this EL mechanism is very similar to the one responsible for EL in forward biased Schottky diodes, namely tunnelling injection of holes from the metal into the n-type semiconductor \cite{Schubert2006b}. The broad band EL in reverse bias can be explained in a similar way as in forward bias.

\begin{figure}[tbp]
   \centering
   \includegraphics[width=0.45\textwidth]{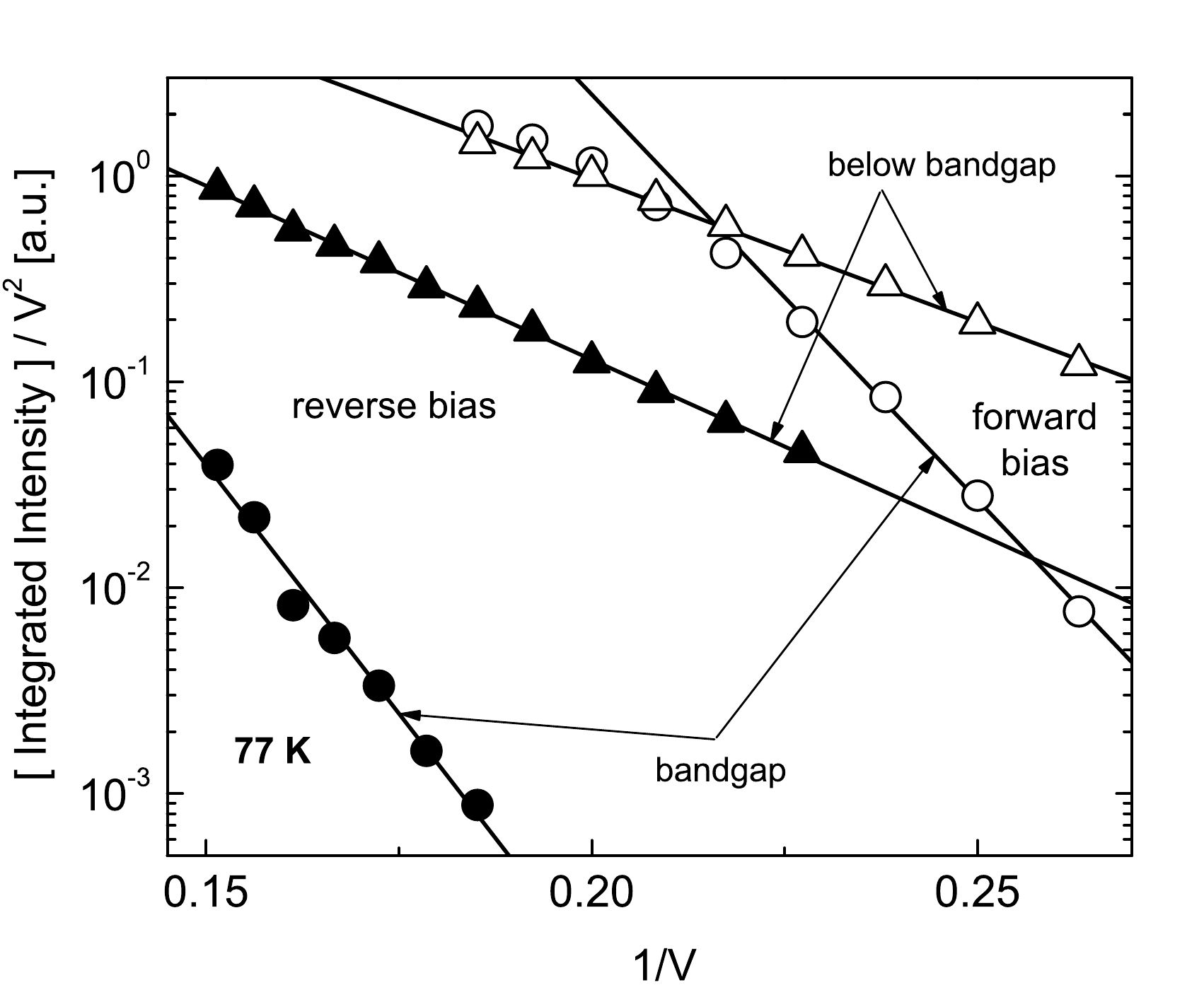}
   \caption{Integrated intensity, divided by $V^2$, versus $1/V$ for bandgap (circles) and sub-bandgap (triangles) EL, under forward (open symbols) and reverse (filled symbols) bias, plotted to exhibit their Fowler-Nordheim-like tunnelling behaviour. The spectral ranges over which the EL was integrated are the same as in the inset of figure \ref{fig:EL}.}
   \label{fig:FN}
\end{figure}

According to the above model, light emission results from radiative recombination of conduction band electrons with holes which first must tunnel through an oxide potential barrier. In the simplest approximation, the tunnel barrier can be modelled as being trapezoidal or triangular in shape, depending on the relative magnitude of the tunnel barrier height $\phi$, and the potential energy difference across the oxide layer $e \eta V$---$\eta$ is the fraction of the applied bias $V$ across the relevant oxide barrier: the nanowire/substrate oxide under forward bias and the nanowire/metal oxide under reverse bias. If $e \eta V < \phi$, the barrier is trapezoidal and the mechanism is referred to as direct tunnelling (DT). On the other hand, when $e \eta V > \phi$, the barrier is triangular and Fowler-Nordheim tunnelling (FNT) takes place \cite{Depas1995}. Within the SimmonÕs approximation \cite{Simmons1963}, we can obtain analytical expressions for both cases. In the limit of small applied bias, i.e. $e \eta V \ll \phi$ , the DT current is proportional to the voltage
\begin{equation}
I \propto V \exp{\left( -\frac{2 d \sqrt{2 m_e \phi}}{\hbar} \right)},
\label{eq:DT}
\end{equation}
where $d$ is the barrier width and $m_e$ is the electron effective mass. At the opposite limit, when $e \eta V$ exceeds the barrier height, the FNT current--voltage dependence can be written as
\begin{equation}
I \propto V^2 \exp{\left( -\frac{4 d \sqrt{2 m_e \phi^3}}{3 \hbar e V} \right)}.
\end{equation}
A plot of $\ln{(I/V^2)}$ against $1/V$ will therefore exhibit logarithmic growth in the DT regime and linear decay in the FNT case \cite{Beebe2006}. It should be emphasized that despite the fact that equation (\ref{eq:DT}) is only exact as $e \eta V \rightarrow 0$, it still holds true that a plot of $\ln{(I/V^2)}$ against $1/V$ will exhibit approximately logarithmic growth in the DT regime even for non-negligible bias across the barrier. Figure \ref{fig:FN} plots the integrated intensities, divided by $V^2$, of bandgap (circles) and below bandgap (triangles) EL, in forward (open symbols) and reverse (filled symbols) bias, as a function of $1/V$ (at 77 K). The figure provides strong evidence for FNT. This result is not surprising from the point of view of the band diagrams in figure \ref{fig:bd}. In the forward bias case, we can provide a quantitative argument supporting this conclusion. For the Si/SiO$_2$ interface, the valence band of the semiconductor is $\sim$4.2 eV below the insulator conduction band, and the bandgap of SiO$_2$ is 8 eV \cite{Shewchun1967}. The barrier for hole tunnelling can then be estimated to be $\sim$3.8 eV. As stated earlier, the potential energy difference across the oxide layer must exceed $\sim$3.5 eV for hole injection to be possible, so that, for the range of applied voltage for which we observe luminescence, we are comfortably in the FNT regime. In reverse bias, however, we cannot provide any quantitative estimates due to lack of knowledge of the band offsets between GaN and its oxide, and the bandgap of the nanowire native oxide. Figure \ref{fig:FN}, however, does highlight the similarity of the EL mechanism in forward and reverse bias. It is also important to recognize that the EL intensities are plotted as a function of the applied bias across the \emph{entire} structure. However, the actual voltage drop responsible for EL is only that part which drops across the nanowire/substrate oxide under forward bias, and across the nanowire/metal oxide under reverse bias. Hence, if the actual voltage could have been used in figure \ref{fig:FN}, the EL intensity under forward bias would likely overlap with the EL intensity under reverse bias, if the two barriers were identical.

Results from all other tested devices were consistent with this interpretation. In particular, we found that \emph{for similar current levels}, devices exhibiting a lower forward bias threshold for near band-edge emission ($\sim$3.4 V) could not be biased significantly under reverse bias ($\sim$4 V), and did not show any appreciable near band-edge luminescence. We interpret this behaviour in terms of a very thin nanowire/metal oxide, which makes it difficult to inject holes into the valence band of GaN from the metal.

The results presented here are of a general nature, in the sense that devices that rely on \emph{mechanical} contacts between two semiconductors will likely behave in a similar way \footnote{For example, in the structure reported in \cite{Duan2003} both the nanowire/substrate and the nanowire/metal interfaces probably have an intermediate oxide layer. As in our case, the nanowire/substrate intervening oxide is probably a combination of the substrate native oxide and the nanowire oxide. However, in contrast to our structure where the nanowire/metal interface is composed only of the nanowire oxide, the structure in \cite{Duan2003} probably has an additional component introduced by the deposition of a 60-80 nm Al$_2$O$_3$ film prior to the metallic contact. The $I\mathrm{-}V$ characteristics reported in the above reference support the presence of an oxide between the nanowire and the substrate. Unfortunately, the reverse bias characteristics were not shown.}. This behaviour stems from the difficulty in making oxide-free semiconductor-semiconductor interfaces \emph{ex situ}. What is more important to note is that if, in fact, it were possible to form a p--n junction between, for instance, the n-GaN nanowire and the p-Si substrate, the large valence band discontinuity ($\sim$2.5 eV) would make it very difficult to inject holes into the GaN. The presence of an oxide barrier is therefore necessary for efficient hole injection to be possible in such structures.

\section{Conclusions}

To summarize, we presented a hybrid nanowire/substrate structure emitting in the ultraviolet under both bias polarities, where current is injected along the length of the nanowire. Tunnel-injection of holes from the p-type Si substrate (under forward bias) and from the metal (under reverse bias) through thin native oxide barriers consistently explains the observed EL behaviour. Our results strongly suggest that the standard p--n junction model is generally not applicable to this kind of device structure.

\begin{acknowledgments}

This work was supported by the National Science Foundation through grant no. ECS-0322720 and by the National Science Foundation Nanoscale Science and Engineering Center (NSEC) under contract NSF/PHY 06-46094. The support of the Center for Nanoscale Systems (CNS) at Harvard University is also gratefully acknowledged. Harvard-CNS is a member of the National Nanotechnology Infrastructure Network (NNIN).

\end{acknowledgments}

\end{document}